\documentclass[aps,pre,showpacs,superscriptaddress,twocolumn]{revtex4-1}
\usepackage{graphicx}
\usepackage{amsmath}
\usepackage{amssymb}
\usepackage{bm}
\usepackage[pdftex]{hyperref}
\hypersetup{colorlinks=true,linkcolor=blue,citecolor=blue,urlcolor=blue}

\begin{document}

\title{Critical Point Scaling of Ising Spin Glasses in a Magnetic Field }

\author{Joonhyun Yeo}\email{jhyeo@konkuk.ac.kr}
\affiliation{Division of Quantum Phases and Devices,
School of Physics, Konkuk University, Seoul 143-701, Korea}
\author{M. A. Moore}\email{m.a.moore@manchester.ac.uk}
\affiliation{School of Physics and Astronomy, University of Manchester,
Manchester M13 9PL, UK}
\date{\today}

\begin{abstract}
Critical point  scaling in a field  $H$ applies for the  limits $t \to
0$,  (where  $t=T/T_c-1$) and  $  H  \to 0$  but  with  the ratio  $R=
t/H^{2/\Delta}$ finite.   $\Delta$ is a critical  exponent of the
zero-field   transition.   We study the replicon correlation length $\xi$ and
from  it the  crossover scaling  function $f(R)$  defined  via $1/(\xi
H^{4/(d+2-\eta)}) \sim f(R)$.   We have calculated analytically $f(R)$
for  the mean-field  limit  of the  Sherrington-Kirkpatrick  model.  In
dimension  $d=3$ we  have determined  the exponents  and  the critical
scaling function  $f(R)$ within two versions of  the Migdal-Kadanoff (MK)
renormalization group procedure.  One of the MK  versions gives
results for  $f(R)$ in  $d=3$  in reasonable  agreement with
those  of the  Monte Carlo  simulations  at the
values  of $R$ for  which they  can be  compared. If  there were  a de
Almeida-Thouless (AT) line  for $d \le 6$ it would appear as a  zero of the function
$f(R)$ at  some negative value  of $R$, but  there is no  evidence for
such behavior. This is consistent with the arguments that there should
be no AT line for $d \le 6$, which we review.

\end{abstract}



\maketitle
\section{Introduction}
\label{sec:Introduction}
Despite decades  of research the nature  of the ordered  state of spin
glasses   remains  controversial.   There  are   two   main  competing
pictures.  The first  is based  on the  broken replica  symmetry (RSB)
ideas of  Parisi and co-workers \cite{Parisi}, which  are indeed exact
in  the limit  of infinite  dimensions --  the Sherrington-Kirkpatrick (SK)
model   \cite{SK}.   The   other   picture  is   the   droplet   model
\cite{Fisher-Huse,Bray-Moore,McMillan},   which   is   based  on   the
properties of excitations (droplets  of reversed spins) in the ordered
phase.

A key discriminator  between the two pictures is  the existence or not
of  the de  Almeida-Thouless (AT)  line \cite{AT}.   According  to the  RSB
picture in  an applied field $H$ there  is a line in  the $H-T$ plane,
$h_{AT}(T)$, at which there  is a transition from the high-temperature
replica symmetric  state to a low-temperature  broken replica symmetry
state. On  the other  hand, in  the droplet picture  there is  no such
transition line.  The  application of the field $H$  removes the phase
transition completely just as does a uniform field applied to an Ising
ferromagnet. There  is no AT  line in the  droplet picture as  the low
temperature  phase in  zero field  is replica  symmetric. It  has been
argued \cite{AT6, Moore12} that  $d=6$ is the lower critical dimension
for the  RSB state and  that there is  no AT line  for $d \le  6$. The
absence  of  RSB for  $d<6$  has  been  rigorously  established  for  a
(unphysical) choice of the distribution function of the spin couplings
$J_{ij}$ \cite{JR}.

 The  question  of  the existence  or  not  of  an  AT line  in  three
 dimensions  can be  investigated experimentally  \cite{Mattsson}. The
 chief  problem  are those of achieving equilibration.   There  are  also
 extensive Monte  Carlo simulations on the properties  of spin glasses
 in  a  field  \cite{JKK,KLY,Larson,Janus1,Janus2}.   For simulations  the  chief
 problem  is that  of  finite size  effects:  These can  give rise  an
 appearance of  RSB even  if it is  absent in the  thermodynamic limit
 \cite{FSBM, Machta}.

In this paper we shall study  critical point scaling.  By this we mean
the  scaling  behavior in  the  limit of  small  fields  close to  the
transition temperature  $T_c$. It is  useful to introduce  the reduced
$t=T/T_c-1$  as  the  measure   of  the  temperature  difference  from
$T_c$. The crossover from the zero field regime to the field dominated
regime    is   measured    by    the    ratio    $R=t/H^{2/\Delta}$
\cite{Fisher-Sompolinsky}. The  exponent $\Delta$ is  the gap exponent
and  is such  that  $\Delta=\beta+\gamma=\nu(d+2-\eta)/2$.  Above  the
upper critical dimension, which is  six for Ising spin glasses, $\beta
= \gamma =1$, $\Delta=2$ and $\nu=1/2$.  (The 
zero-field  correlation length varies as  $1/t^{\nu}$).  The decay  of the
zero-field     bond-averaged    spin-spin     correlation    function,
$\overline{\langle S_i  S_j \rangle^2}$ with  distance at $T_c$  is as
$1/r^{d-2+\eta}$; $\eta=0$  for $d>6$. There  is the usual   failing of
hyperscaling for  $d>6$. For  $d <6$,  all the
exponents take on non-trivial values and have been extensively studied
\cite{Janus3}   in  three  dimensions   and  investigated  via  epsilon
expansions  below six  dimensions  \cite{Harris-Lubensky-Chen,YMA}. In
the presence  of a  field there are  many correlation lengths  and the
longest of these  is the replicon length scale  $\xi$ \cite{AT}. Right
at $T_c$, this grows as  $\xi \sim 1/H^{4/(d+2-\eta)}$ and for $d>6$
this reduces  to $\xi \sim  1/H^{1/2}$. When $t$ is  non-zero, $\xi$
becomes
\begin{equation}
\frac{1}{\xi H^{4/(d+2 -\eta)}} \sim f \left(\frac{t}{H^{2/\Delta}}\right)=f(R).
\label{fdef}
\end{equation}
The focus  of this  paper is on  the form  of the function  $f(R)$. We
shall refer  to it as  the crossover function  as it describes  how the
correlation  length at $T=T_c$  (i.e. $t=0$)  is  modified  when $t$  is
non-zero. It is important to  keep in mind that Eq.  (\ref{fdef}) only
strictly applies  in the limits $t \to  0$, $H \to 0$,  with the ratio
$R$ fixed. For applications one always  has $t$ and $H$ finite and one
needs to  allow for  corrections to scaling.  We can  determine $f(R)$
analytically in the  mean-field limit, i.e. for the  SK model and this
is  done in  Sec. \ref{sec:SKlimit}.   For three  dimensions  we shall
calculate  it with two  different versions  of the  Migdal-Kadanoff RG
procedure in Sec. \ref{sec:MK&Janus} and compare them with the results
of the Monte Carlo simulations of Ref. \cite{Janus2}.

Since there is  an AT line for $d  >6$ and in the SK model  we need to
understand  why there  is no  trace of  it in  the  crossover function
$f(R)$. The answer  lies in the scaling form of the  AT line, i.e. the
behavior  of $h_{AT}$  as  $|t|  \to 0$.   When  $d>8$ $h_{AT}^2  \sim
|t|^3$,    but    for    $8>d>6$,    $h_{AT}^2    \sim    |t|^{d/2-1}$
\cite{AT,Fisher-Sompolinsky,Green-Moore-Bray}.   Then for all  $d >6$,
$R$  goes to  $-\infty$  at the  AT line  in  the limit  $|t| \to  0$.
However, for $d<6$, if there is an AT line, it would occur at a finite
value     of     $-R$      as     $h_{AT}^2     \sim     |t|^{\Delta}$
\cite{Fisher-Sompolinsky}.   Because $\xi$  diverges at  the  AT line,
$f(R)$ has to  have a zero if  there is an AT line.  However, we shall
see no evidence for such a zero in the work for $d=3$ reported in Sec.
\ref{sec:MK&Janus}.   The  absence  of  an  AT  line  for  $d  =3$  is
consistent with the argument of Ref.  \cite{AT6} that the AT line only
exists  when  $d   >  6$.    This   argument  is   sketched  in   Sec.
\ref{sec:ATline}: It  is because  for $8>d  >6$,  $h_{AT}^2 \sim(d-6)^4
|t|^{d/2-1}$, {\it when $|t| \to  0$}, which indicates that 
in the  scaling limit,  the AT  line is going  away as  $d$ approaches
$6$. The comments of Ref. \cite{PT} on this argument will be reviewed.

\section{The crossover function $f(R)$ in the SK limit}
\label{sec:SKlimit}
The Hamiltonian  for the SK Ising  spin glass in a  field \cite{SK} is
given by
\begin{equation}
\mathcal{H}=-\sum_{\langle ij\rangle}
J_{ij} S_i S_j
-H \sum_i  S_i,
\label{SKHam}
\end{equation}
where  the  Ising spins  $S_i$  take  the  value $\pm1$  and  $\langle
ij\rangle$ means  that the  sum is  over all pairs  $i$ and  $j$.  The
couplings   $J_{ij}$  are   chosen  independently   from   a  Gaussian
distribution  with   zero  mean  and  a   standard  deviation  (width)
$J/N^{1/2}$.  We  shall calculate  for  this  Hamiltonian the  scaling
function $f(R)$. This is easy to  do as the calculation is done in the
region  where there  is  replica symmetry.  There the  Edwards-Anderson
order parameter $q=1/N \sum_i \langle S_i \rangle^2$ is obtained by solving the equation
\begin{equation}
q=\int_{-\infty}^{\infty}\,\frac{dx}{\sqrt{2 \pi}}e^{-x^2/2}
 {\rm tanh}^2(\beta J \sqrt{q}x+\beta H).
\label{qSK}
\end{equation}
It is convenient to introduce the notation $Q=\beta^2J^2q+\beta^2H^2$,
 and $t=T/T_c-1$, where for the SK model $T_c=J$. Then to order $t^2$ and $h^2 (=H^2/T_c^2)$, one can obtain by expanding the argument of  tanh in Eq. (\ref{qSK}) the following approximation to the equation of state
\begin{equation}
H^2/T_c^2 \equiv h^2=2 t Q+2 Q^2.
\label{eqnstate}
\end{equation}

To determine the crossover function $f(R)$, we need to determine the replicon correlation length. This is the length scale associated with the decay of the (bond-averaged) replicon correlation function $G_R(ij)$,
\begin{equation}
G_R(ij)=\overline{(\langle S_i S_j\rangle-\langle S_i \rangle \langle S_j \rangle)^2}.
\label{GRdef}
\end{equation}
For the replica symmetric state, de Almeida and Thouless \cite{AT}
calculated the eigenvalues of the Hessian associated with the replica symmetric solution. The smallest eigenvalue was that in the replicon sector
and is given by
\begin{equation}
\lambda_R=1-\beta^2 J^2 \int_{-\infty}^{\infty}\,\frac{dx}{\sqrt{2 \pi}}e^{-x^2/2}
{\rm sech}^4(\beta J \sqrt{q}x+\beta H).
\end{equation}
To order $t$ and $h$ this reduces to $\lambda_R=2t+2 Q$. Eliminating $Q$ using Eq. (\ref{eqnstate}) gives 
\begin{equation}
\lambda_R \approx t+\sqrt{t^2+2 h^2}.
\label{lambdaapprox}
\end{equation}

We next use the Ornstein-Zernike approximation to determine the replicon correlation length by setting $1/\xi^2 \sim \lambda_R$. Then we get on using the notation $R=t/h$, which is what $R=t/h^{2/\Delta}$ becomes for $\Delta=2$,
\begin{equation} 
\frac{1}{\xi h^{1/2}}=f(R) \sim (R+\sqrt{2+R^2})^{1/2}.
\label{f(R)SK}
\end{equation}
Note that the power of  $H$ in Eq. (\ref{fdef}), $4/(d+2-\eta)$, can be
re-written as $2 \nu/\Delta$, which reduces  to $1/2$ for the SK model where
 $\nu  =1/2$, and  $\Delta  =2$.   Eq.  (\ref{f(R)SK})  is  the
mean-field approximation  for $f(R)$. Exactly the same  result for the
crossover function  $f(R)$ is obtained for a  Gaussian distribution of
fields  whose variance is  $H^2$, so  there are  universality features
associated with it.  In Sec.  \ref{sec:MK&Janus} the form of $f(R)$ is
studied  in  three  dimensions  and   it  is  found  to  be  at  least
qualitatively   similar   to  that   of   the   mean-field  limit   in
Eq. (\ref{f(R)SK}).

Notice that when $t$ is negative $f(R)$ decreases to zero as $R \to -\infty$.
\begin{equation}
f(R) \sim 1/|R|^x, \hspace{0.5cm} R \to -\infty,
\label{xdef}
\end{equation} 
where   in  the   SK  limit   the   exponent  $x$   takes  the   value
$\frac{1}{2}$. We  would expect  the same value  for $x$ to  hold from
infinite dimension down to the upper critical dimension, $d=6$.

\section{The AT line for $d>6$}
\label{sec:ATline}
Since  we are  essentially  studying the  replicon correlation  length
$\xi$, one might  have expected to see in  the scaling function $f(R)$
behavior associated with  the AT line. The AT line is  the line in the
$H-T$ plane at which $\xi$ goes  to infinity.  Near $T_c$, the AT line
for the SK model is given  by $h^2_{AT} \sim |t|^3$ \cite{AT}, so on the
AT line $R = t/H^{2/\Delta} = t/h_{AT}$. This diverges to $-\infty$ in
the critical  scaling limit $|t| \to  0$: The AT line  is thus outside
the critical scaling limit for the SK model, which explains its absence
from $f(R)$.

We next review how the AT line evolves with dimension $d$. The SK form
for the AT line as $|t| \to  0$ is expected to hold down to $d=8$, and
changes    for    $8>d>6$,     to    $h_{AT}^2    \sim    |t|^{d/2-1}$
\cite{Green-Moore-Bray,  Fisher-Sompolinsky}. Provided  $d>6$,  $R$ at
the AT line is outside the  critical scaling limit. However, as $d \to
6^+$, the  value of $R$ on the  AT line is diverging  to infinity less
and less  strongly (as $1/|t|^{\frac{d-6}{4}}$). A  change must occur
at $d=6$.  {\it If there were an  AT line for $d\le 6$, it would be in
  the critical scaling region and occur at a finite value of $R$}.  In
Sec.  \ref{sec:MK&Janus}  we shall  study the computer  simulations of
the Janus group  \cite{Janus2} and the MK approximation  for $f(R)$ in
three dimensions for evidence for such a zero at a finite value of $R$
and  find none.   The  more detailed  calculation  of Ref.  \cite{AT6}
provide a  clue as to what  happens: The AT transition  does not occur
for $d \le 6$.

The   more   detailed  calculations   started   from  the   replicated
Ginzburg-Landau-Wilson free-energy functional for the Ising spin glass
which,  written  in  terms   of  the  replica  order  parameter  field
$Q_{\alpha \beta}$, is
\begin{eqnarray}
&F[\{Q_{\alpha\beta}\}]  =   \int   d^dx\,   \left[\frac{1}{2}r
\sum_{\alpha<\beta}Q_{\alpha\beta}^2
+\frac{1}{2}\sum_{\alpha<\beta}(\nabla
Q_{\alpha\beta})^2\right. \nonumber  \\
&\hspace*{-0.2cm}  +   \left.\frac{w}{6}
\sum_{\alpha<\beta<\gamma}Q_{\alpha\beta}
Q_{\beta\gamma}Q_{\gamma\alpha}  -  h^2
\sum_{\alpha<\beta} Q_{\alpha\beta} + O(Q^4).\right]\nonumber\\
\label{FQ}
\end{eqnarray}
The coefficient  $r$ is essentially a
measure of the  distance from $T_c$ i.e. it is  basically $t$. When $d
<8$  the $Q^4$  terms  are irrelevant.  Conventional  RG methods  were
applied \cite{Harris-Lubensky-Chen},  but the calculation was done 
above the upper  critical dimension, $d_u=6$. It is useful to
define $\epsilon = d-6$. Then for small $\epsilon$ \cite{AT6,PT}
\begin{equation}
h^2_{AT} \sim \frac{w|r|^{d/2-1}}{\left[(2w^2/\epsilon)(1-|r|^{\epsilon/2})+1\right]^{5d/6-1}}.
\label{match}
\end{equation}
In Moore and Bray \cite{AT6} it was argued that Eq. ({\ref{match}) implied
that near six dimensions in the limit when $|r| \to 0$,
\begin{equation}
 h_{AT}^2 \sim (\frac{\epsilon}{2w^2})^4 w |r|^{d/2-1}.
\label{MB6}
\end{equation}
This  result strongly  suggests that the AT  line will
disappear as $d \to 6$ in the  critical  scaling limit,
 $|r| \to  0$.   At fixed
$|r|$ in the limit of $\epsilon \to 0$, Eq. (\ref{match}) gives
\begin{equation}
h_{AT}^2 \sim (\frac{1}{w^2 \ln |r|+1})^4 w |r|^{d/2-1},
\label{PTAT}
\end{equation} 
which agrees with the expression for the AT line in six dimensions given by Parisi and Temesv\'{a}ri \cite{PT}. However, Eq.~\eqref{match} is not valid for this limit.

To see that, it is useful to note that the general form of the AT line, (at least for $6 <d <8 $), is 
\begin{equation}
h_{AT}^2 =\frac{|r|^2}{w} g(w^2 |r|^{\epsilon/2}),
\label{PTform}
\end{equation}
At small  values  of  $y=w^2 |r|^{\epsilon/2}$,
one    can   construct   the    perturbative   expansion    for  $g(y)$
\cite{Green-Moore-Bray}.  The  renormalization group calculation which
leads  to Eq. (\ref{match}) is  effectively just  a resummation  of the
perturbative calculation  and will only  be useful and valid  at small
values of $y$, which means only for the critical scaling limit  $|r| \to 0$ at 
fixed $\epsilon$.

However, we  would agree  with the authors  of Ref. \cite{PT}  that it
would be  good to have  an argument that  at any fixed value  of $|t|$,
$h_{AT}$  went to zero in the  limit $d \to 6$,  rather than just
for the  scaling limit $|t| \to  0$. This is  a non-perturbative task,
but perhaps not impossible.  In Ref.  \cite{Moore12} a $1/m$ expansion
of  the value  of  the AT  field,  $H_{AT}$, at  zero temperature  was
undertaken, for an  $m$-spin component spin glass ($m=1$  is the Ising
spin glass,  $m=3$ is the Heisenberg  spin glass).  It  was found that
the first term in the $1/m$ expansion went to zero around $d=6$.

\section{Droplet scaling and  $f(R)$ for $d \le 6$}
\label{sec:droplet}

In Sec. \ref{sec:MK&Janus} we find that there is no evidence that there is an AT line 
in $d=3$. In other words, the crossover function $f(R)$ has no finite zero on the negative
axis. In the limit when $R \to -\infty$,  it decays towards zero, as $1/|R|^{x}$.
In this section we show that this behavior for the  crossover function  is 
predicted by droplet scaling and determine how the exponent $x$ depends on the
critical exponents and $\theta$.

According to the droplet picture  the length scale $\xi$ is determined
by an Imry-Ma  argument \cite{Fisher-Huse,Bray-Moore} where the energy
cost  of  flipping  a  region  of  linear  extent  $\xi$,  $\Upsilon
\xi^{\theta}$,  is balanced  against the  magnetic field  energy which
could  be gained, $\sqrt{q}  H \xi^{d/2}$.  The interface  energy term
$\Upsilon$   scales  as   $|t|^{\nu   \theta}$  and   $q$  scales   as
$|t|^{\beta} \sim |t|^{\nu(d-2  +\eta)/2}$  at  small  $|t|$.  Thus  according  to  the
droplet picture
\begin{equation}
\frac{1}{\xi} \sim \left [\frac{\sqrt{q}}{\Upsilon} H \right ]^{\frac{1}{d/2-\theta}} \sim \left [ \frac{H}{|t|^{\nu[\theta-(d-2+\eta)/4]}} \right]^{\frac{1}{d/2-\theta}}.
\label{Imry-Ma}
\end{equation}
The  Imry-Ma argument  for $\xi$  is a  scaling argument  itself which
should hold for all $T< T_c$ in  the limit $H \to 0$. As a consequence
it should  coincide with  the $R \to  -\infty$ limit for  the critical
scaling function,  at least for $d  \le 6$. This enables  us to relate
the exponent  $x$ for the  decay of $f(R)$  as $1/|R|^x$ to  the other
exponents.   Multiplying  both   sides  of   Eq.   (\ref{Imry-Ma})  by
$1/H^{\frac{4}{d+2-\eta}}$ gives
\begin{equation}
\frac{1}{\xi H^{4/(d+2-\eta)}} =  \left [ \frac{1}{|t|^{\nu[\theta-(d-2+\eta)/4]}} \right]^{\frac{1}{d/2-\theta}}H^{(\frac{1}{d/2-\theta}-\frac{4}{d+2 -\eta})}.\\
\label{largeRscaling}
\end{equation}
This will  approach $\frac{1}{|R|^x}$ as $R \to -\infty$. Note that  
\begin{equation}
\frac{1}{|R|^x}=\left [\frac{H^{\frac{4}{\nu(d+2-\eta)}}} {|t|}\right ]^x.
\label{consistency}
\end{equation}
The exponents of both $|t|$ and $H$ in Eqs. (\ref{consistency}) and (\ref{largeRscaling}) must both agree. This happens if
\begin{equation}
x= \frac{\nu}{d/2-\theta}\left[\theta-\frac{d-2+\eta}{4} \right].
\label{xfix}
\end{equation}

We shall  give numerical estimates  of the value  of $x$ for  $d=3$ in
Sec. \ref{sec:MK&Janus}. For $d>6$,  $x=\frac{1}{2}$. We do not expect
that Eq.   (\ref{xfix}) should necessarily give  $x=\frac{1}{2}$ as $d
\to 6^{-}$.   This is  because for $d>6$,  $x$ is associated  with the
Gaussian  perturbative  fixed  point,   while  for  $d\le  6$,  it  is
associated with  the zero-temperature  fixed point of  droplet scaling
(see  Sec.  \ref{sec:discussion} for  a  further  discussion of  fixed
points).

\section{The crossover function $f(R)$ in three dimensions}
\label{sec:MK&Janus}

In  this section  we shall  study two  variants (called  Scheme  I and
Scheme  II) of  the  Migdal-Kadanoff  RG procedure  to  first get  the
critical exponent  $\eta$ and then  the crossover function  $f(R)$. We
find the Janus  result for $\eta$ lies between the  values of Scheme I
and that of  Scheme II.  The MK results for  $f(R)$ have been obtained
over  a  much  wider range  of  $R$  than  those  of the  Janus  group
\cite{Janus1, Janus2, Janus3}.  For Scheme II there is reasonable agreement  on $f(R)$ with that of the Janus group at the $R$ values at which they 
they can be directly  compared. Scheme I is less satisfactory; it also leads to a value for $\eta >0$ and a value for $x<0$, both of which seem rather unlikely. There is no good evidence of a zero of $f(R)$  at
a finite value of $R$ in either Scheme, which means they provide no evidence for an AT line
in three dimensions.

The Edwards-Anderson Hamiltonian for the Ising spin glass in a field is given by
\begin{equation}
\mathcal{H}=-\sum_{\langle ij\rangle}
J_{ij} S_i S_j
-H \sum_i  S_i,
\end{equation}
where the Ising spins $S_i$ take the value $\pm1$ and $\langle ij\rangle$ means that the sum is over all nearest neighbor pairs $i$ and $j$. 
The couplings $J_{ij}$ are chosen independently from a Gaussian distribution with zero mean 
and a standard deviation (width) $J$.  
The temperature $T$ enters into the problem through $\exp(-\mathcal{H}/T)$ and the flows of the couplings $J_{ij}/T$ 
and the fields $H/T$ are studied in the Migdal-Kadanoff RG. 

We use the  approximate bond moving schemes.  For  a three dimensional
cubic lattice, $2^{d-1}=4$ (when $d=3$) bonds are put together to form
a new  bond. The coupling constant of  this new bond is  just given by
the  sum  of the  coupling  constants of  the  four  bonds. Using  the
notations in Fig.~\ref{bond_move}, we have
\begin{equation}
 J^\prime_1=\sum_{i=1}^4 J_1^{(i)},~~~~~ J^\prime_2=\sum_{i=1}^4 J_2^{(i)}.
\end{equation}

As for the external field, however, 
there is a certain freedom in how the on-site field variables are moved in the bond-moving scheme.
In this paper, we employ two different schemes for moving the fields. 
The first one, which we call Scheme I, is due to Ref.~\cite{DBM}. In this scheme,
when three bonds are moved to 
combine with a bond, the fields on the three bonds are moved to the
site that is to be traced over. This procedure prevents the fields from increasing indefinitely, and allows us
to work with the uniform applied field.
In terms of the notations in Fig.~\ref{bond_move}, we have
\begin{equation}
  H^\prime_1=H_1^{(1)},~~~~~~H^\prime_2=H_4^{(1)}, 
  \end{equation}
  and
  \begin{equation}
  H^\prime_3=\sum_{i=2}^4 H_1^{(i)}+\sum_{i=1}^4H_2^{(i)}+\sum_{i=1}^4H_3^{(i)}+\sum_{i=2}^4 H_4^{(i)}.
 \end{equation}

In the other scheme, Scheme II, the fields stay with the bonds when moved. This kind of field moving method was used 
in Ref.~\cite{Cao} for the random field Ising model. From Fig.~\ref{bond_move}, we assign in this case
\begin{equation}
  H^\prime_1=\sum_{i=1}^4 H_1^{(i)},~~~~~
  H^\prime_2=\sum_{i=1}^4 H_4^{(i)}, 
  \end{equation}
  and
  \begin{equation}
  H^\prime_3=\sum_{i=1}^4 H_2^{(i)}+\sum_{i=1}^4H_3^{(i)}.
 \end{equation}
Since the field grows indefinitely as the  iteration continues, the uniform
external  field is not  appropriate in  this case.   Instead we  use a
random external field of zero mean and the standard deviation $H$.  In
both   schemes,  once  the   renormalized  fields,   $H^\prime_n$  are
determined, a trace is performed over the spins at the site connecting
the two new bonds.  The decimation procedure can be continued $n$ times
for a system of size $L=2^n$.

\begin{figure}
 \includegraphics[width=\columnwidth]{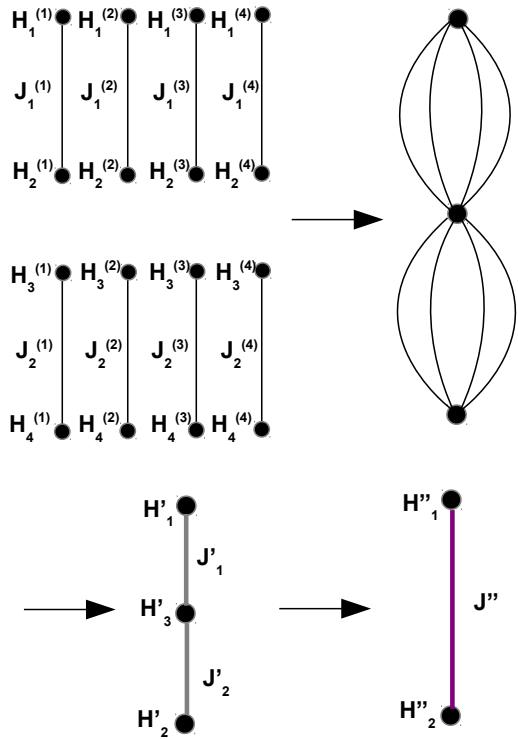}
 \caption{(Color  online) Bond-moving  scheme  in the  Migdal-Kadanoff
   approximation. In the  first step two sets of  four bonds are moved
   together.  The  renormalized couplings  and fields are  assigned in
   the second step according to  the given scheme. Finally, a trace is
   taken over the middle spin.}
 \label{bond_move}
\end{figure}

We perform the  MK RG numerically for given  temperature $T$ and given
uniform field $H$ (Scheme I) or given field width $H$ (Scheme II). All
these  quantities are  measured in  units  of $J$.   In the  numerical
calculations, we prepare $10^6$ bonds.  On each bond the couplings are
chosen independently from the Gaussian distribution with zero mean and
width  $1/T$.   On  each end  of  the  given  bond, we  either  assign
$H/(2dT)$  for  Scheme  I  or  choose  the  field  from  the  Gaussian
distribution of zero mean and  width $H/(2dT)$. The factor of $1/(2d)$
is used to account for  the coordination number in the $d$-dimensional
cubic lattice.   We then randomly  select two sets  of 4 bonds  out of
$10^6$ to  form two new bonds,  and follow the  procedure described in
Fig.~\ref{bond_move} to obtain renormalized couplings and fields. This
procedure is continued until we  get $10^6$ new bonds, which completes
the first iteration.  As we iterate the same  procedure, we can obtain
the flow of the couplings and fields as a function of iteration number
$n$.

\begin{figure}
 \includegraphics[width=\columnwidth]{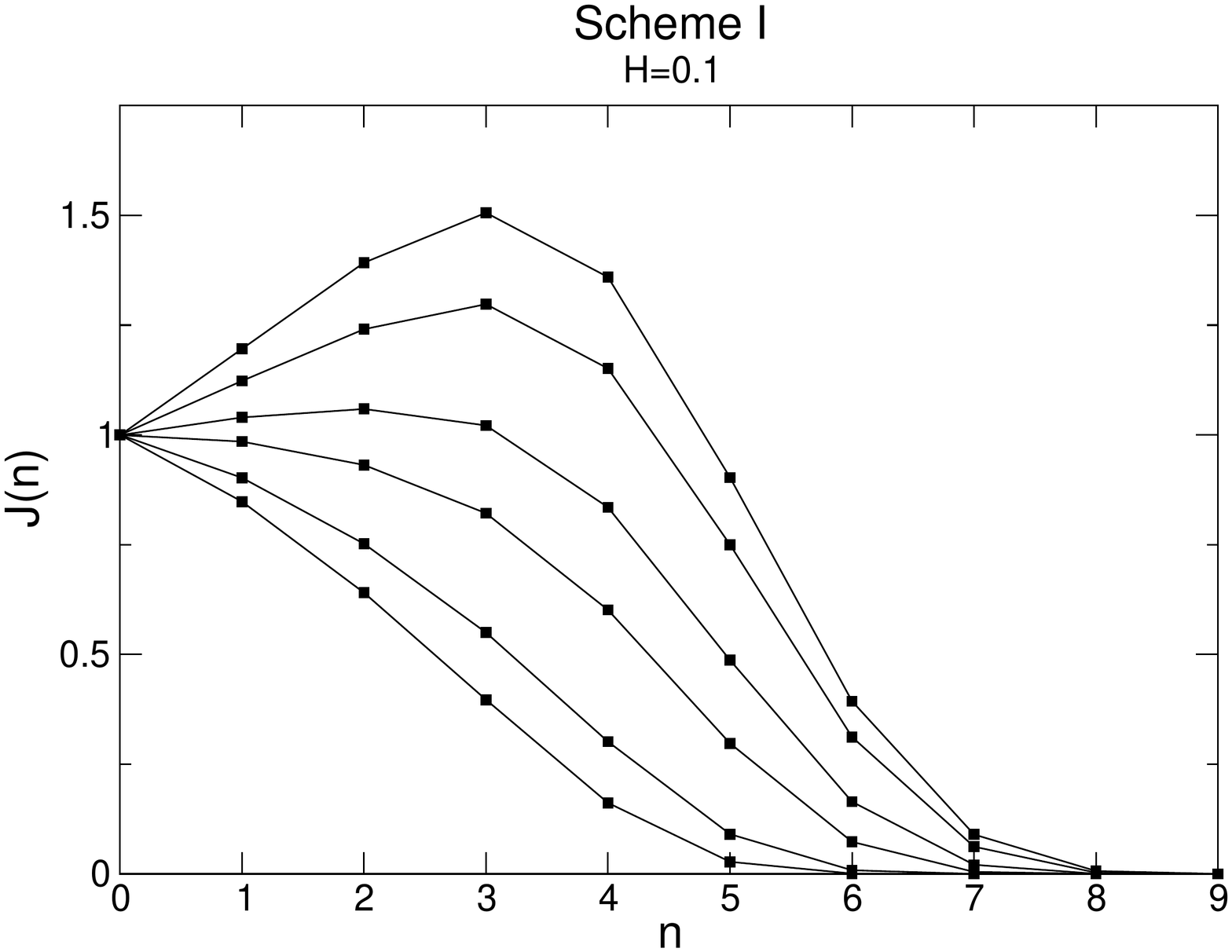}
 \includegraphics[width=\columnwidth]{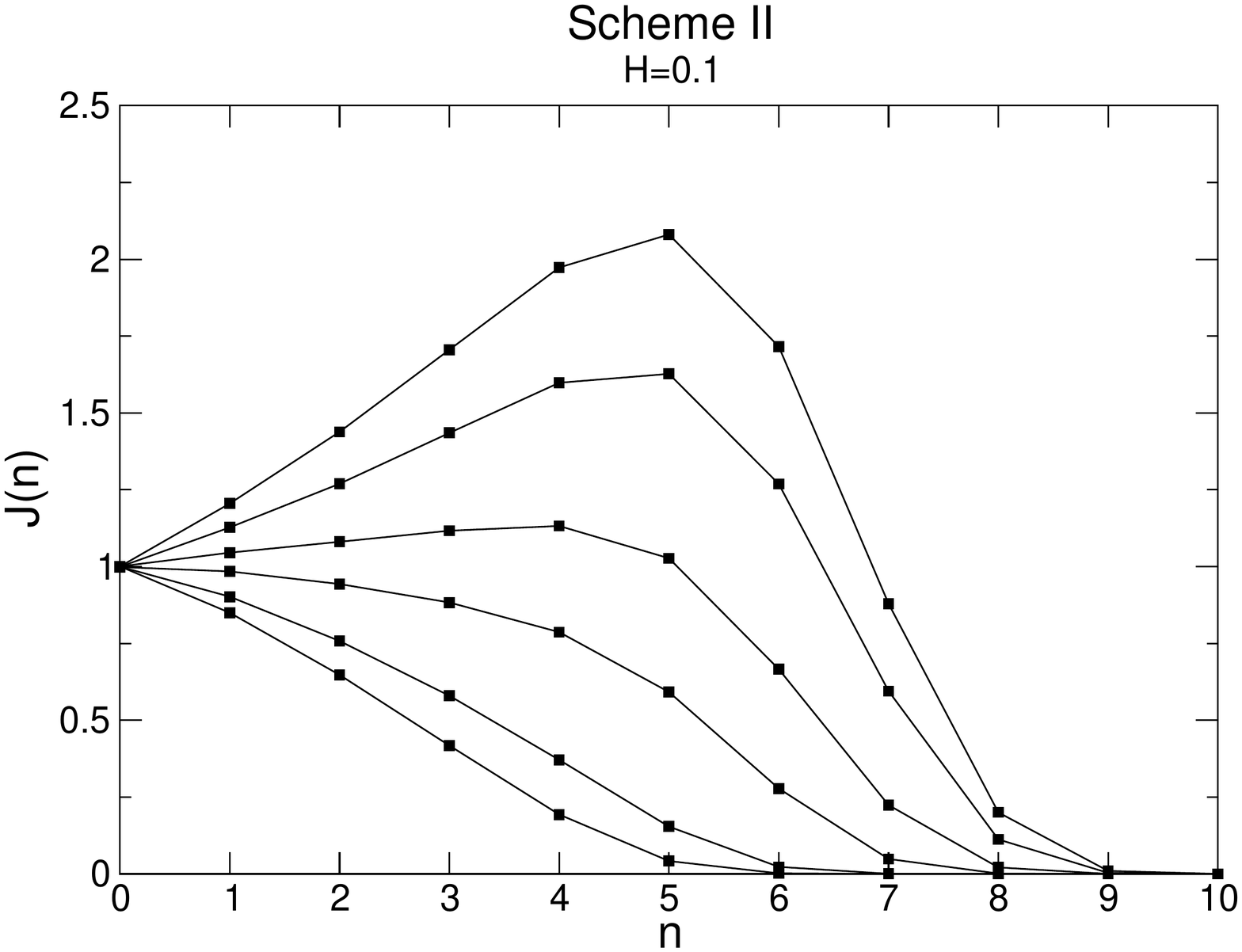}
 \caption{The coupling strength $J(n)$  at iteration step $n$ at fixed
   external field  $H=0.1$ in Scheme  I (upper figure) and  Scheme II,
   (lower figure).   The temperatures are $T=0.1$, 1.0,  1.6, 2.0, 2.6
   and 3.0 from top to bottom }
 \label{coupling_s}
\end{figure}

At each  step of the iteration  we measure the  standard deviations of
the couplings, $J(n)$. The RG flow of these widths of the couplings at
fixed   external  field  is   shown  in   Fig.~\ref{coupling_s}.    For a
finite field, the  coupling strength always flow to  zero in all cases
we studied indicating the absence  of a phase transition.  In general,
the decay of the coupling  strength is slower at low temperatures. The
decaying part  of $J(n)$ can be described  as $\exp[-L/\xi(T,H)]$ with
$L=2^n$, where the $\xi(T,H)$ is interpreted as the correlation length
at temperature $T$ and field $H$. Note that the decay in the Scheme II
is slower than  that in Scheme I, so the  correlation length in Scheme
II is generally larger than that in Scheme I.

\begin{figure}
 \includegraphics[width=\columnwidth]{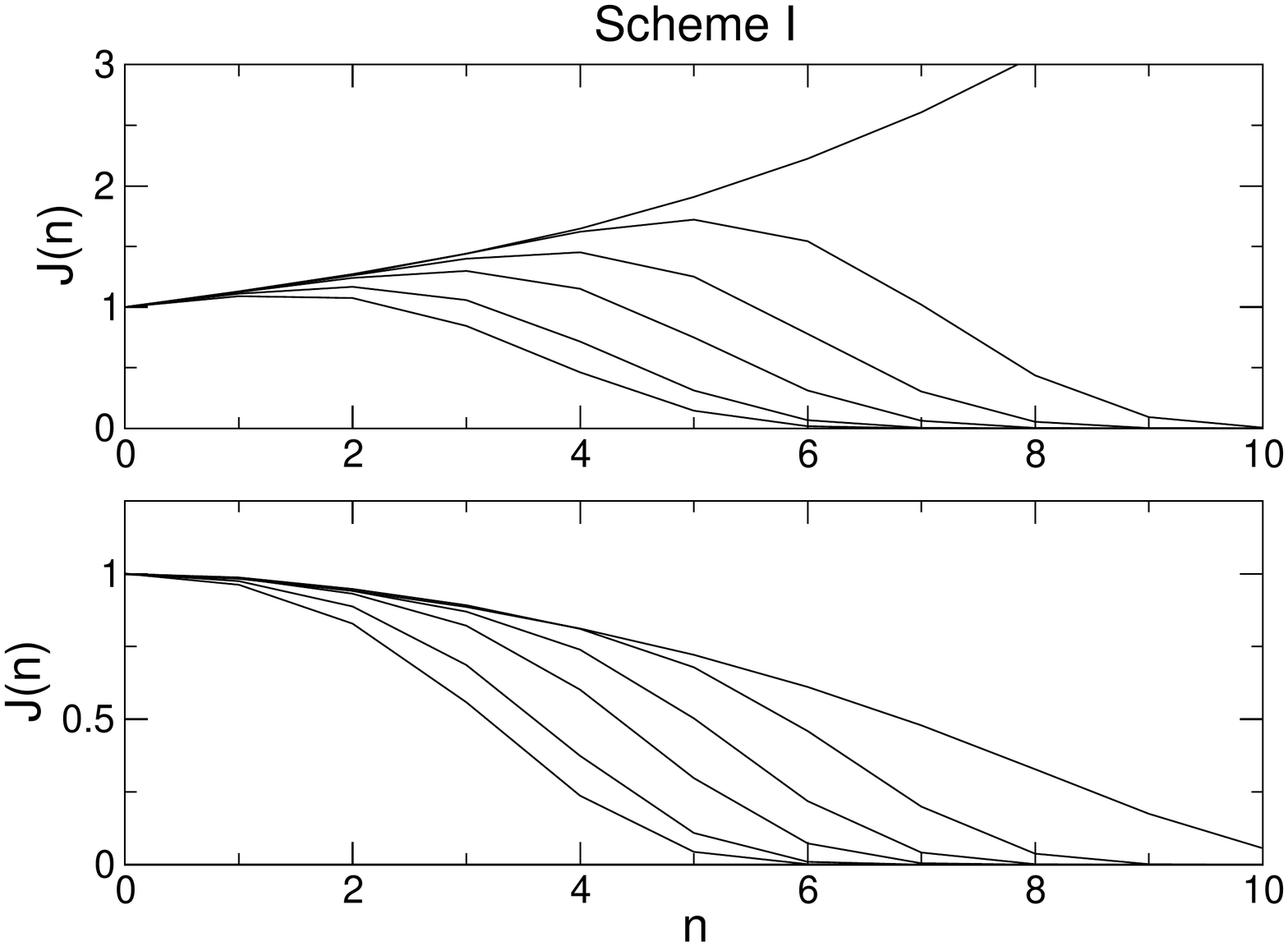}
 \includegraphics[width=\columnwidth]{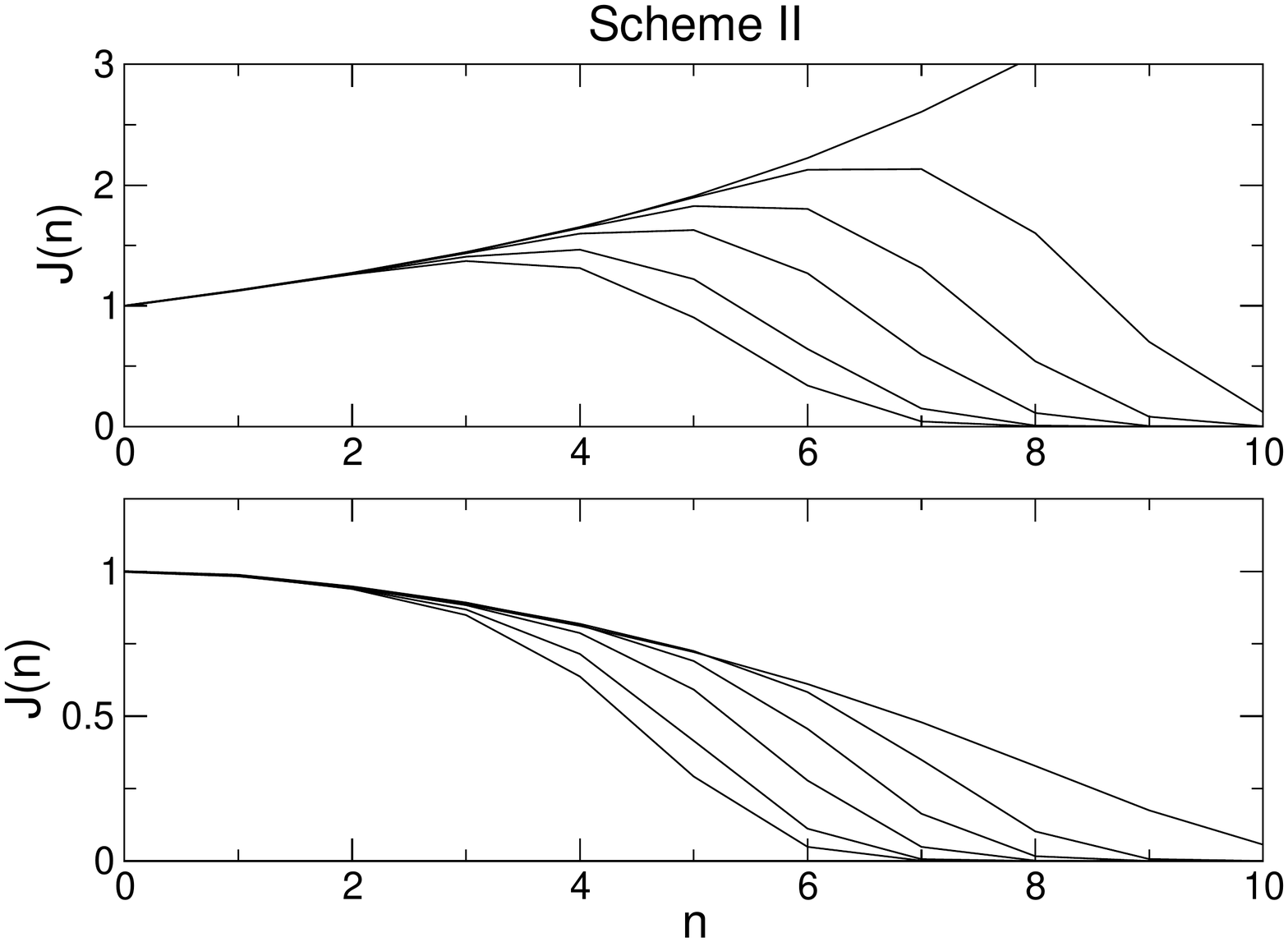}
 \caption{The flow of the coupling $J(n)$ at fixed temperature and varying fields in Schemes I and II . The upper panel for each Scheme is at $T=1.0$ and
 the lower one is for $T=2.0$. From top to bottom, $H=0$, 0.02, 0.05, 0.1, 0.2, and 0.3.}
 \label{coupling_h_s}
\end{figure}

In Fig.~\ref{coupling_h_s}, the RG flow of the couplings is shown at fixed temperature. 
The results at zero field are also included in these figures. We can see that the coupling strength flows to infinity if $T<T_c$ for $H=0$.
We estimate the zero-field transition temperature as $T_c=1.77$.  
Note that the presence of the external field makes the correlation length decrease.  
\begin{figure}
 \includegraphics[width=\columnwidth]{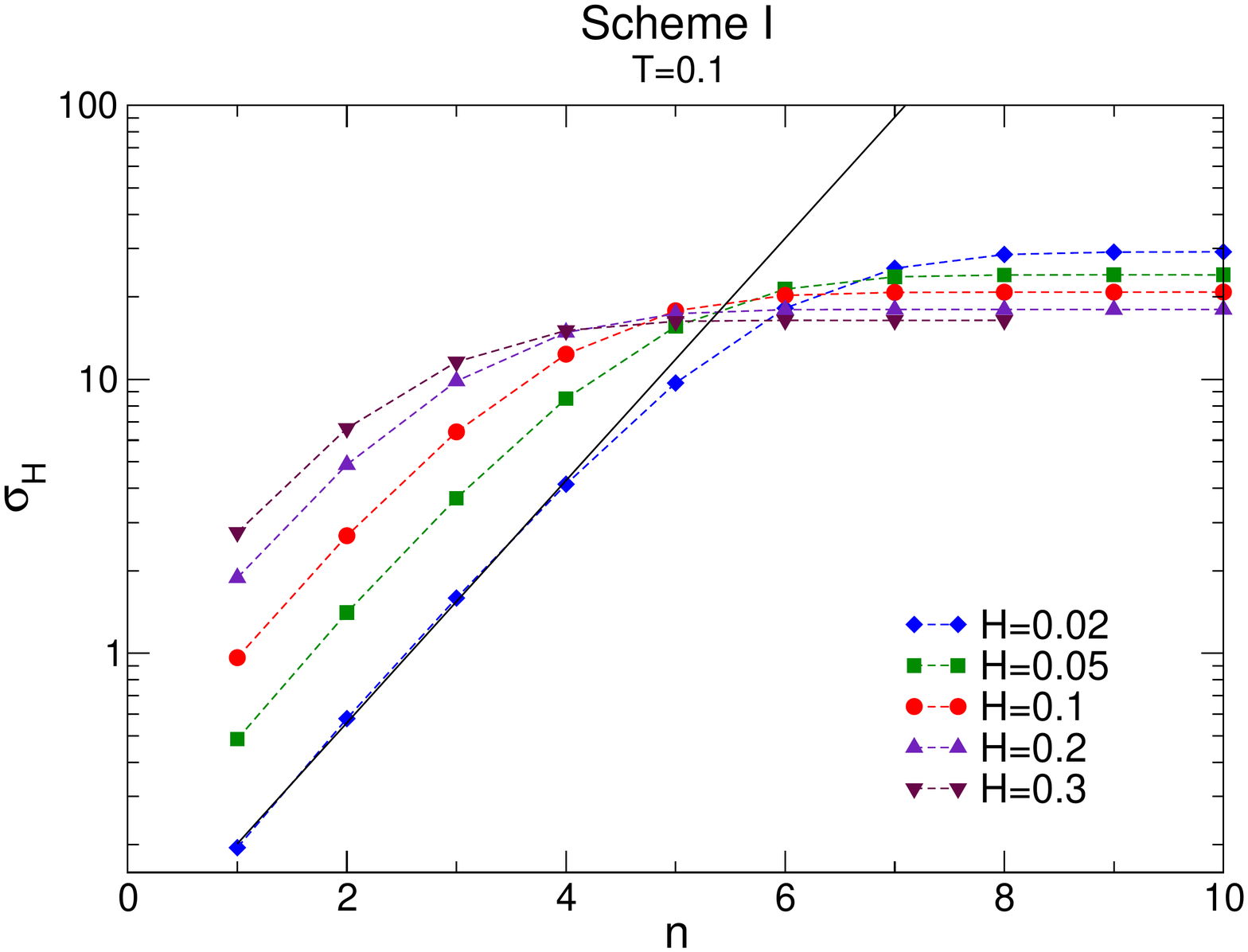}
 \includegraphics[width=\columnwidth]{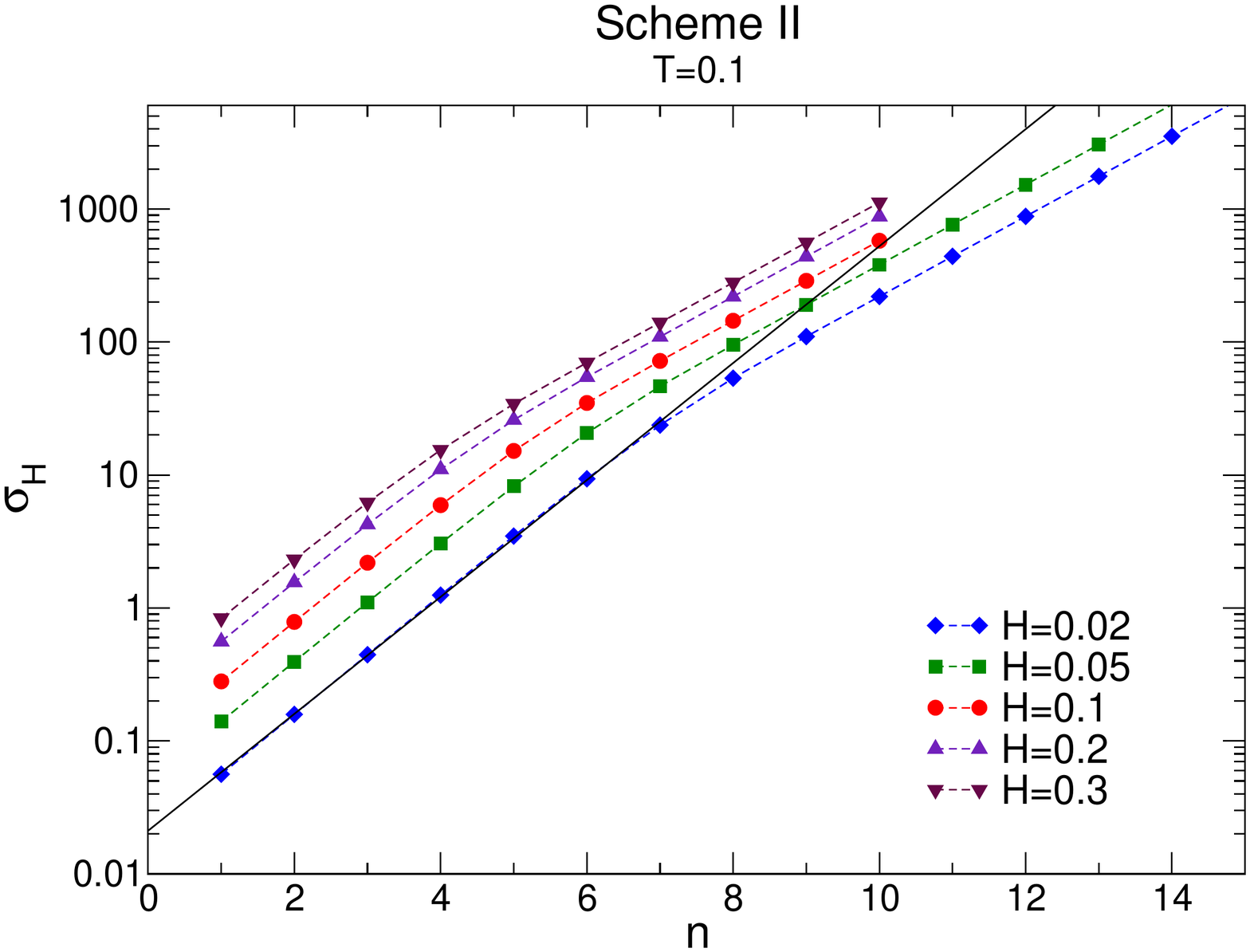}
 \caption{(Color online) The standard deviation of the fields as a function of iteration in Scheme I (upper figure)  for fixed temperature $T=0.1$ and varying 
initial uniform field $H$. The solid line in the upper figure is the fit $\sigma_H\sim L^{1.47}$. The lower figure is for Scheme II also for fixed temperature  $T=0.1$ and varying standard deviation $H$ of the initial random field. The solid line is a  fit  to   $\sigma_H\sim L^{1.46}$. }
 \label{field_width_s}
\end{figure}

\begin{figure}
 \includegraphics[width=\columnwidth]{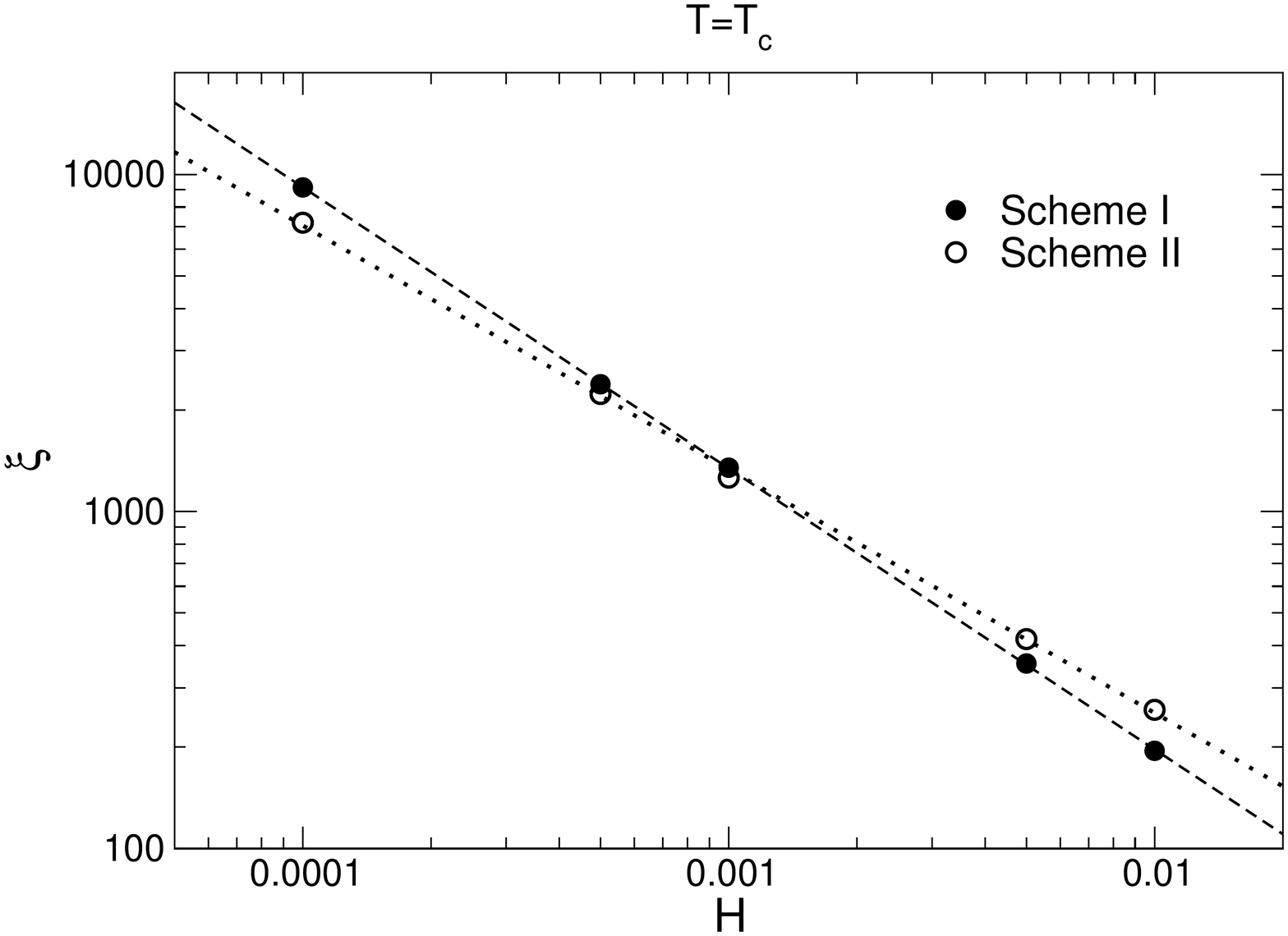}
 \caption{The correlation length $\xi(T_c,H)$ at zero-field transition
   temperature $T_c$  for small external field $H$.   The dashed lines
   are fits, $\xi\sim H^{-0.83}$  for Scheme I and $\xi\sim H^{-0.72}$
   for Scheme II.}
 \label{xi_h_tc}
\end{figure}

In  Scheme  I,  we start  from  the  uniform  external field.  As  the
iteration continues, however, the  field becomes no longer uniform but
shows  a random distribution.   The mean  value of  the fields  is not
changed, but the standard  deviation $\sigma_H$ keeps increasing until
it saturates after the coupling drops  to a small value as can be seen
in  Fig.~\ref{field_width_s}.  In Scheme  II,  the standard  deviation
$\sigma_H$ also  increases from the initial  value $\sigma_H(0)=H$ for
the random external field  as shown in the lower figure of
  Fig.~\ref{field_width_s}.  But
it does not saturate in this case, but increases at a slower rate.  In
both cases,  the initial increase  of $\sigma_H$ is well  described by
$\sigma_H\sim L^{d/2}$ as  $T\to 0$.

Finally we  use our  results for the  critical exponents and  $\xi$ to
determine the  crossover function $f(R)$. The results  are displayed in
Fig. \ref{crit_inv_s1} for Scheme  I and in Fig. \ref{crit_inv_s2} for
Scheme  II.
\begin{figure}[ht]
 \includegraphics[width=\columnwidth]{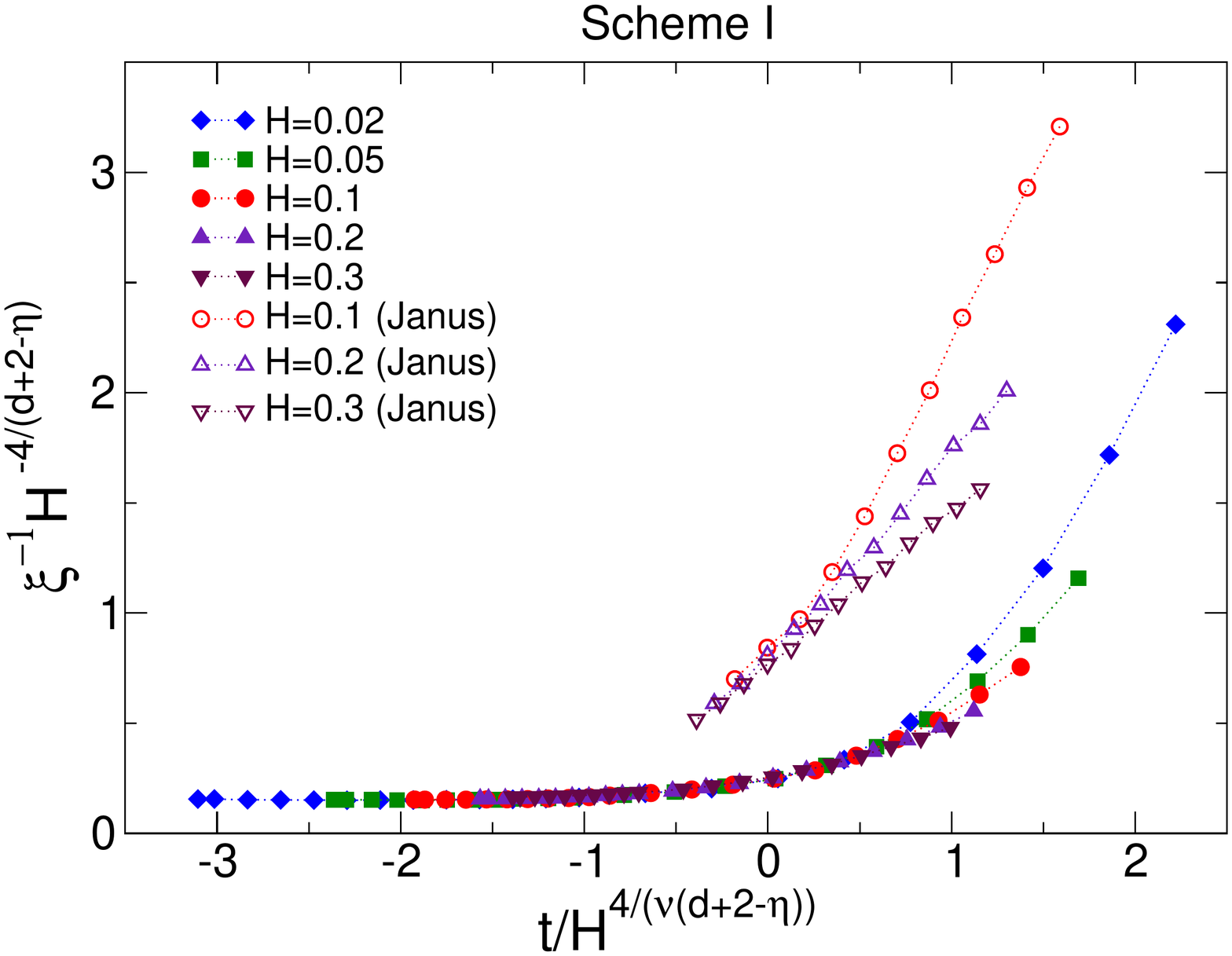}
 \includegraphics[width=\columnwidth]{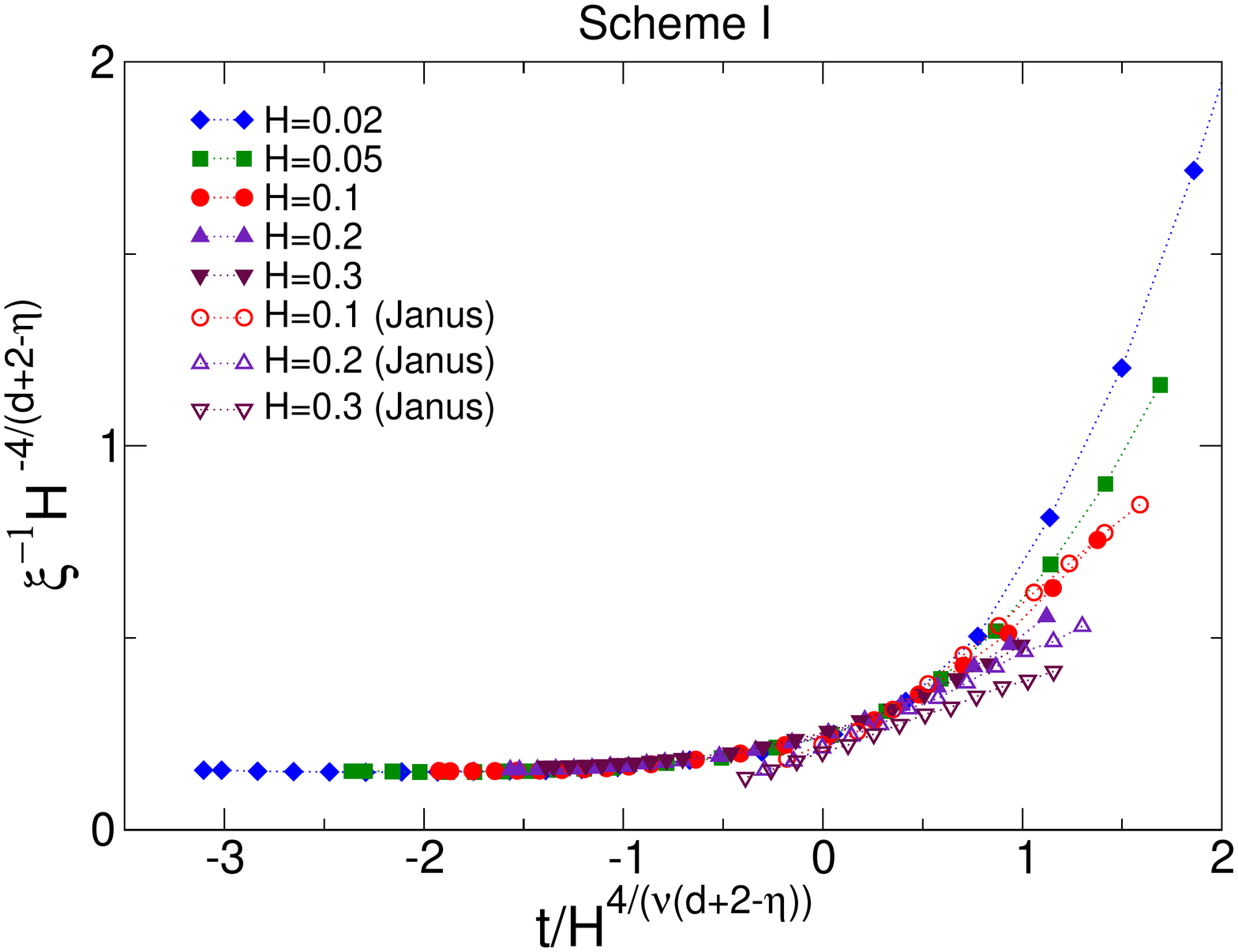}
 \caption{(Color online) The upper panel is  $\xi^{-1} H^{\frac{4}{d+2-\eta}} \equiv f(R)$ versus $t/H^{4/(\nu(d+2-\eta)}=R$ 
 for MK Scheme I and Janus data \cite{Janus2}. 
 The lower panel is the same but with  the Janus data for $\xi$ rescaled by the factor 3.79, as described in the text. 
 }
 \label{crit_inv_s1}
\end{figure}

\begin{figure}[ht]
 \includegraphics[width=\columnwidth]{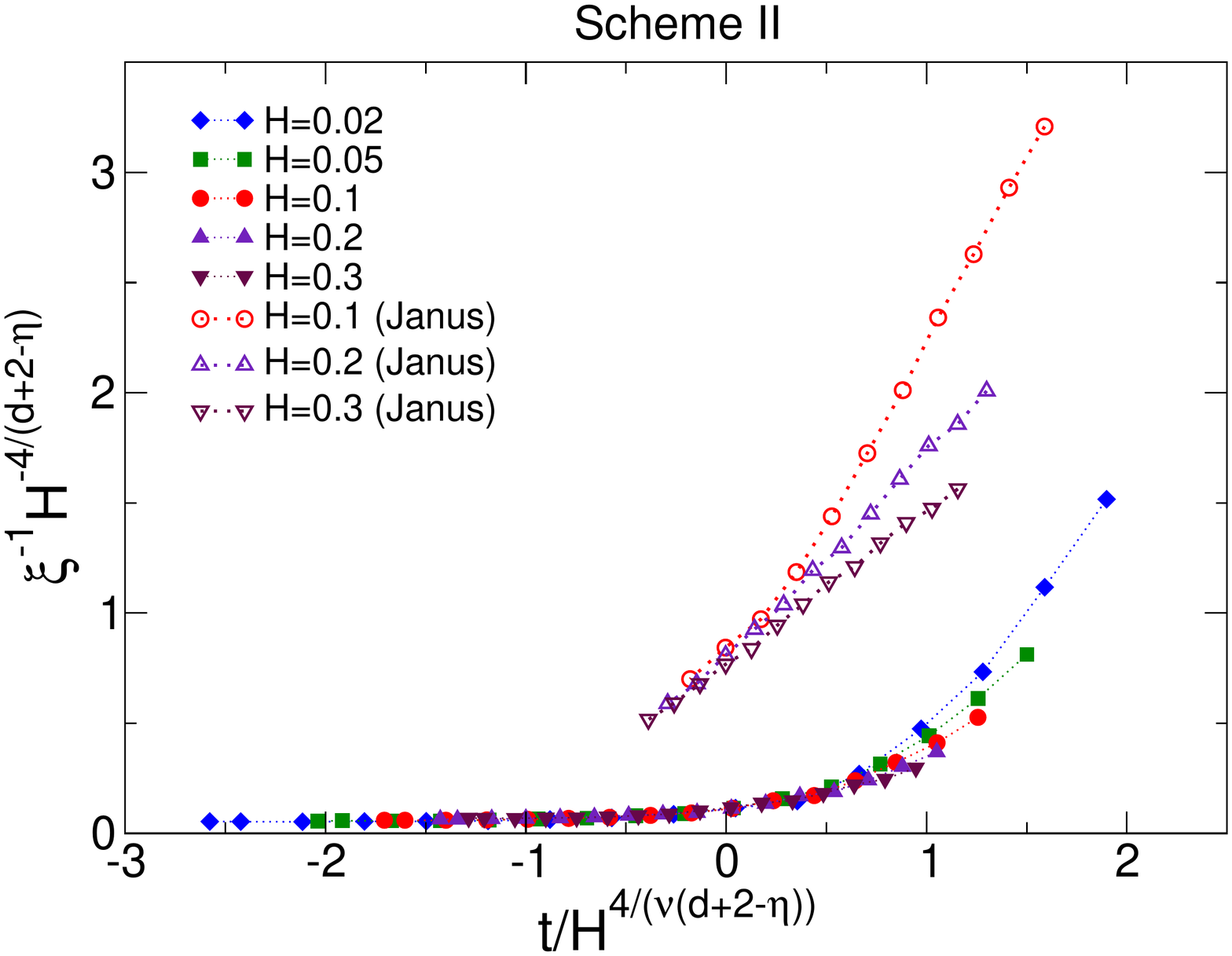}
 \includegraphics[width=\columnwidth]{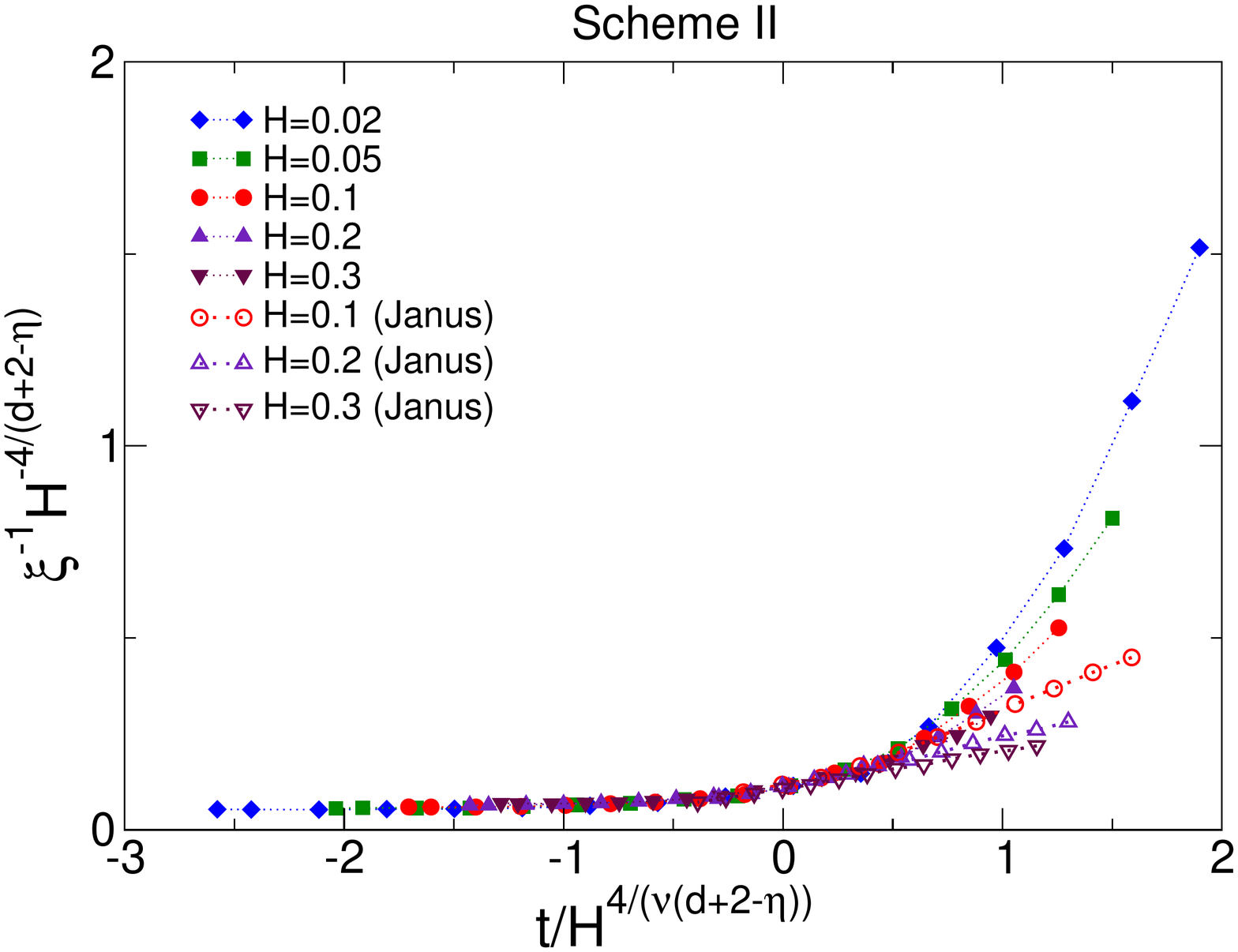}
 \caption{(Color online) The upper panel is  $\xi^{-1} H^{\frac{4}{d+2-\eta}} \equiv f(R)$ versus $t/H^{4/(\nu(d+2-\eta)}=R$ 
 for MK Scheme II and Janus data \cite{Janus2}. The lower panel is the same but with  the Janus data for $\xi$ rescaled  
 by the factor 7.13, as described in the text.}
 \label{crit_inv_s2}
\end{figure}
 Also   on  the  plots  are  the   results  of  the  Janus
collaboration \cite{Janus2}.  In neither case is  there good agreement
between the MK results and the Janus simulation. However, the apparent
discrepancy is just  an artifact due to using  different definitions of
$\xi$. We extracted  $\xi$ by studying the decay  of the couplings $J$
with iteration number.
The Janus  $\xi$ was obtained from a particular
moment of the replicon  correlation function $G_R$.
 A different choice
of moment  would alter their  estimate of $\xi$. However,  one expects
the different  definitions of $\xi$  to have essentially the  same $t$
and  $H$  dependence, i.e.  they  differ  mostly  by a  multiplicative
constant.      We      chose      that      constant      to      make
$\xi_{Janus}(t=0,H=0.2)=\xi_{MK}(t=0,H=0.2)$.   For   Scheme  I   that
constant is 3.79 while for Scheme II it is 7.13. 
We then multiplied the Janus data by that factor and the resulting plots are as shown in the lower panels of Figs.  \ref{crit_inv_s1} and  \ref{crit_inv_s2}. The Janus data is now in reasonable agreement with the MK results over the ranges of $R$ for which there is overlapping scaling data at least for Scheme II. However, it is only for the MK results that we have data over a large enough range to see the decay of $f(R)$ as $1/|R|^x$ at large values of $-R$.
There is no evidence that $f(R)$ has a zero at any finite negative value of $R$, which implies that there is no AT line. 

We end this Section by summarizing the values found or used for the various exponents.\\

{\it  Critical point scaling exponents}:\\
  (1) The     exponent     $\eta$     was     evaluated     from
     Fig.~\ref{xi_h_tc}.  From $\xi(T_c,H)\sim  H^{-4/(d+2-\eta)}$, we
     have $\eta\sim 0.18$ for Scheme I and $\eta\sim -0.56$ for Scheme
     II. The Janus collaboration \cite{Janus3} estimate is $\eta \sim =-0.39$.\\
     (2) The  MK  value   for   $1/\nu=0.356$  
     \cite{Southern} for both  Schemes  I and
     II. The Janus estimate \cite{Janus3} is  $1/\nu=0.39$.

{\it Droplet Scaling Exponents}:\\ (1) We found by studying the growth
of $J(L)\sim L^\theta$ at zero  field and at very low temperature that
$\theta=0.26$ .  The value obtained for $d=3$ in Ref. \cite{Boettcher}
was $\theta=0.24$.\\
(2) Using Eq.  (\ref{xfix}) in $d=3$ for MK scheme I,  $x = -0.079$, for Scheme  II, $ x=0.34$. For the Janus exponents, $x
= 0.18$.

\begin{figure}[ht]
 \includegraphics[width=\columnwidth]{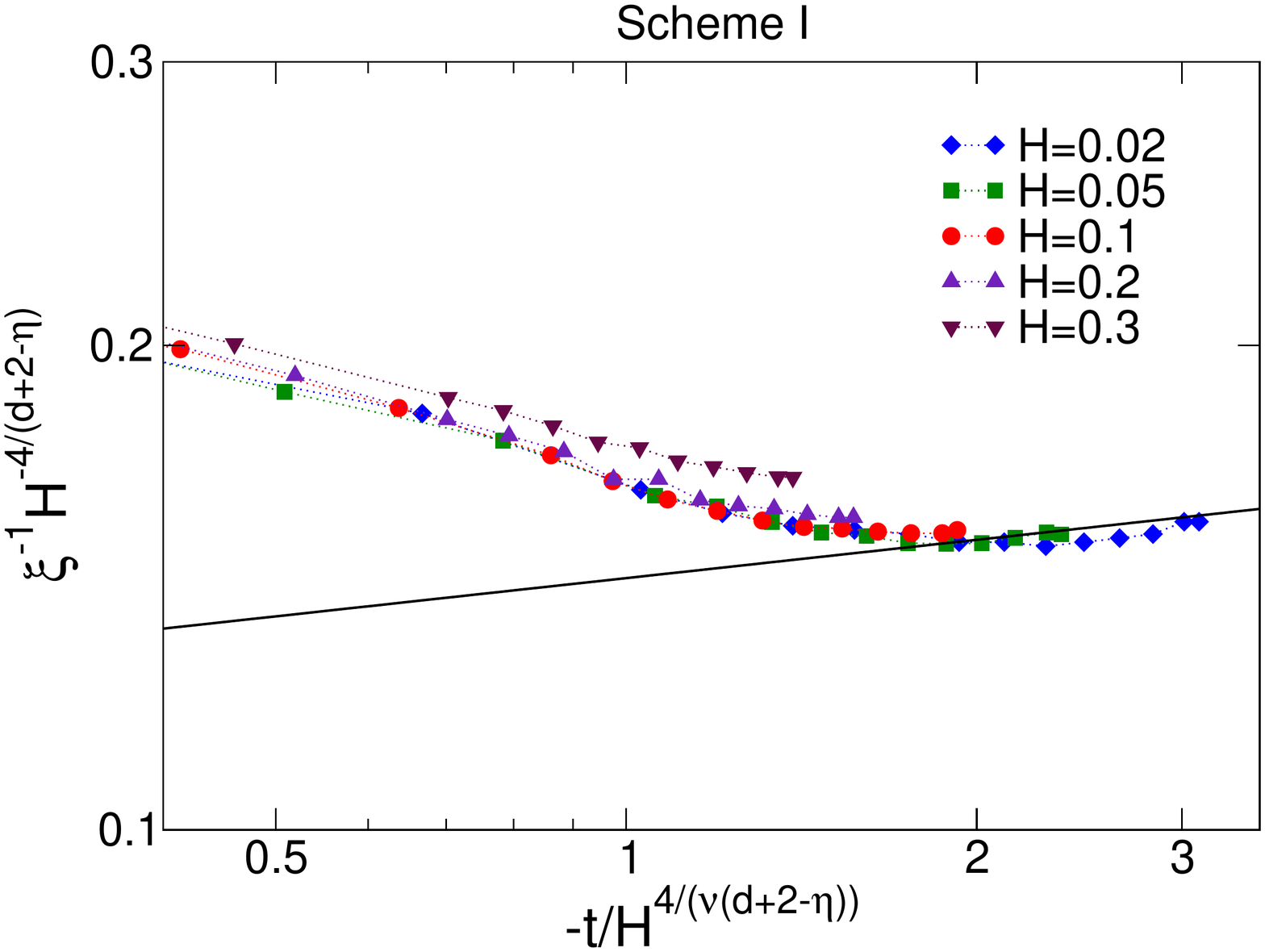}
 \includegraphics[width=\columnwidth]{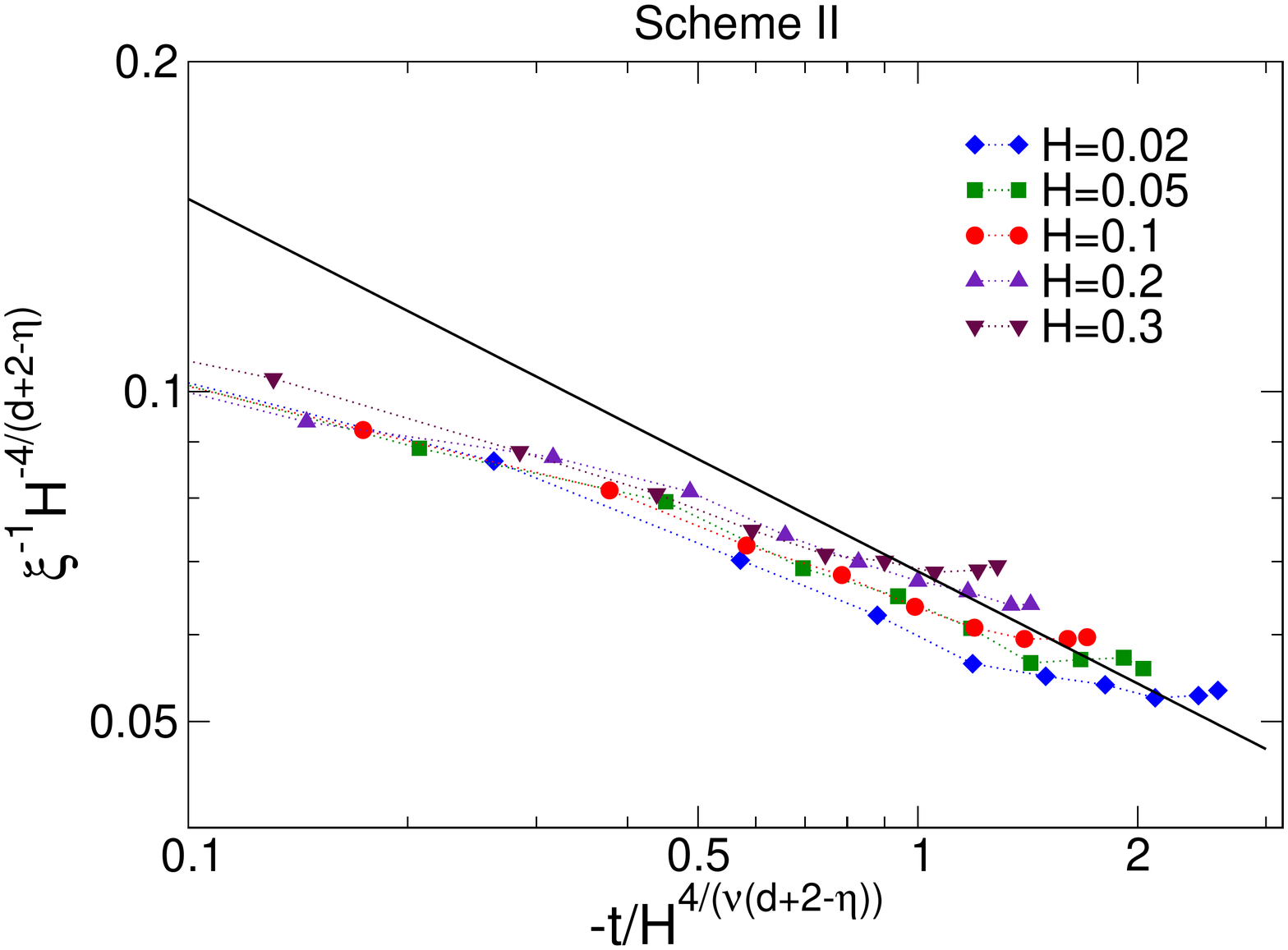}
 \caption{(Color online) The log-log plot of $\xi^{-1} H^{\frac{4}{d+2-\eta}} \equiv f(R)$ versus $-t/H^{4/(\nu(d+2-\eta)}=-R$ 
 for MK Scheme I (upper panel) and Scheme II (lower panel)
 to illustrate the power law behavior $f(R) \sim 1/|R|^x$ 
 which is predicted to set in at large values of $-R$. The straight line is the expected slope if $x$ is predicted by 
 Eq. (\ref{xfix}). 
 }
 \label{loglog}
\end{figure}

\section{Discussion}
\label{sec:discussion}

In
Figs. \ref{crit_inv_s1} and \ref{crit_inv_s2}
the MK results include data points  right down to $T=0$ which are well
outside the  critical regime.   Some of the  points from low  $T$ have
departures from  the universal curve,  which are visible if  the large
$|R|$ region is put on an  expanded scale, as in Fig.~\ref{loglog}. The reason for this is that
the critical scaling  forms  $\Upsilon  \sim  |t|^{\nu \theta}$  and  $q  \sim
|t|^{\beta}$ do not hold accurately  right down to $T=0$.  But despite
this, the critical scaling collapse of the data is surprisingly good. 

Critical scaling strictly applies only  for the limit $t \to 0$, $H\to
0$, with the ratio  $R=t/H^{2/\Delta}$ fixed. One always hopes though
that there  is a sizable critical  region where the  scaling forms are
good  approximations. We  have  seen that  when  $T<T_c$ the  critical
region extends  to a  good approximation down  to $T=0$.  However, the
figures  show   the  critical  region   is  much  smaller  when
$T>T_c$.  Many of the MK and Janus data points are  off the scaling
line, indicating that they were  calculated from values of $t$ and $H$
which were too large to be in the scaling region.

Another  surprise is that  when $d<6$,  the critical  scaling function
$f(R)$ involves  the   exponent  $\theta$,  which  is  an   exponent  of  the
zero-temperature fixed point. The RG flows near the AT line were first
investigated by Bray and Roberts \cite{Bray-Roberts} in a perturbative
treatment appropriate for $d \to  6^{-}$ and no stable fixed point was
found. It has been suggested from time to time that such runaway flows
perhaps indicate that the behavior is controlled by a zero temperature
fixed point. Within the MK  approximation it is possible to follow the
flows  of  the  fields  $H$   and  the  couplings  $J$  under  the  RG
iterations. If one  starts out at a temperature  below $T_c$, when the
fields $H$ are  very small, the flows take one very  close to the zero
temperature fixed point associated  with $H=0$ before they finally run
off to  large $H/J$ values. Thus it  may be that the  RG runaway flows
found by  Bray and Roberts  are related to  the effects produced  by a
zero-temperature  fixed point.   A role  for a  zero-temperature fixed
point  has  also  been  suggested  recently  by  Angelini  and  Biroli
\cite{Angelini}   in   a   very   unconventional  scenario   for   the
high-dimensional behavior of spin glasses in a field.

\acknowledgements We  should like to  that V. Martin-Mayor,  G. Parisi
and A.P. Young for useful discussions and observations. JY was supported by
Basic Science Research Program through the National Research Foundation 
of Korea (NRF) funded by the Ministry of Education (NRF-2014R1A1A2053362).

\end{document}